\documentstyle[aps,psfig,amssymb]{revtex}
\begin{document}
\title{Damage of cellular material under simultaneous application of
pressure and pulsed electric field}
\author{
M. I. Bazhal$^{a,b}$ ,
N. I. Lebovka$^{a,c}$ 
, E. Vorobiev$^{a} $\footnote{Corresponding author,
E-mail:Eugene.Vorobiev@utc.fr}}
\address{
$^{a}$D\'{e}partement de G\'{e}nie Chimique, Universit\'{e} de
Technologie de Compi\`{e}gne, Centre de Recherche de Royallieu,
B.P. 20529-60205 Compi\`{e}gne Cedex, France
\\ $^{b}$ Ukrainian State University of Food Technologies, 68, Volodymyrska str.,
Kyiv, 252033, Ukraine
\\ $^{c}$ Institute of Biocolloidal
Chemistry named after F.D. Ovcharenko, NAS of Ukraine, 42,
blvr.Vernadskogo, Kyiv, 252142, Ukraine }
\maketitle
\begin{abstract}
\renewcommand{\baselinestretch}{0.2}
\parskip=3mm
\tighten

Influence of pulsed electric field (PEF) simultaneous to pressure
treatment on moisture expression from fine-cut cellular raw material has been
investigated. Dependencies of specific conductivity $\sigma$,
liquid yield $Y$, instantaneous flow rate $v$ and qualitative juice
characteristics at different modes of PEF treatment are discussed.
Three main consolidation phases were
observed in a case of mechanical expression. A unified approach is
proposed for liquid yield data analysis allowing
to reduce the data scattering caused by differences in the quality
of samples. Simultaneous application of pressure and PEF treatment
allowed to reveal a passive form of electrical damage.
Pressure provokes the damage of defected cells,
 enhances diffusion migration of moisture in porous cellular material
and depresses the cell resealing processes.
PEF application at a moment when a sample specific electrical
conductivity reaches minimum and pressure achieves its constant
value seemed to be the most optimal.
\end{abstract}
\pacs{
\renewcommand{\baselinestretch}{0.2}
\parskip=3mm
\tighten Keywords: Cellular material; Permeabilization; Plasmolysis; Pressing; Pulsed electric field
treatment
 }



\begin{tabular}{ll}
{\bf Notation} &
\\ $d$ & mean cell's dimension, $\mu m$
\\ $D$ & diffusion coefficient, m$^{2}$ s$^{-1}$
\\ $E$ & electric field strength, kV cm$^{-1}$
\\ $k$ & area normalizing coefficient
\\ $M^{1}$ & first Moment, s
\\ $N$ & number of pulses
\\ $P$ & pressure, bar
\\ $S$ & area under the expression curve $Y(t)$, \% s
\\$t_{i}$ & pulse duration, $\mu $s
\\$t_{max}$ & maximal time of pressing, s
\\$t_{p}$ & time of PEF treatment application, s
\\$t_{v_{max}}$ & time, where maximum of instantaneous flow rate is observed, s
\\ $\Delta t$ & pulse repetition time, ms
\\ $T$ & temperature, K
\\ $v$ & $=dY/dt$ instantaneous flow rate, \% s$^{-1}$
\\ $U$ & external voltage, V
\\ $W$ & moisture content, \%
\\ $Y^{I}$ & first normalized form of liquid yield
\\ $Y^{II}$ & second normalized form of liquid yield
\\ $Y ^{*}$ & intensification degree of PEF treatment

\end{tabular}

\begin{tabular}{ll}
{\it Greek letters}
\\ $\sigma $ & conductivity, S m$^{-1}$
\\ $\tau $ & characteristic time of expression, s
\\ $\tau ^{*}$ &  coefficient of PEF enhanced durability
\\ $\tau_{D}$ & $\sim d^{2}/6D$  time constant of diffusion process, s
\end{tabular}

\begin{tabular}{ll}
{\it Subscripts}
\\$E$ & with PEF treatment
\\$E=0$ & without PEF treatment
\\$r$ & reduced to the maximal value
\\$\infty$ & in the limit of infinite time

\end{tabular}

\begin{tabular}{ll}
{\it Abbreviations}
\\PEF & pulsed electric field
\\SEM & scanning electron microscopy
\end{tabular}

\section{Introduction}

Mechanical expression (hydraulic pressing) is widely used in the
processes of solid-liquid separation for extraction of fruit
juices and vegetable oils, dewatering of fibrous materials, etc.
(Schwartzberg, 1983). Efficiency of this process
can be increased by raw material plasmolysis, cellular damage or
permeabilization prior to its expression. Different methods are
traditionally used to increase the degree of raw material
plasmolysis: heating, osmotic drying or freezing dehydration,
alkaline breakage, enzymatic treatment, etc.
(Rao \& Lund, 1986; Aguilera \& Stanley, 1999; Tsuruta, Ishimoto
\& Masuoka, 1998; Ponant, Foissac \& Esnault, 1988; Jones, 1988;
Barbosa-C\'{a}novas \& Vega-Mercado, 1996). Earlier on, the method
of electric field treatment (both d.c. and a.c.) was also proposed
for cellular material plasmolysis (known as electro-plasmolysis).
The methods of electro-plasmolysis were shown to be good for juice
yield intensification and for improving the product quality in
juice production (Scheglov, Koval, Fuser, Zargarian, Srimbov,
Belik et al., 1988; McLellan, Kime \& Lind, 1991, Bazhal \&
Vorobiev, 2000), processing of vegetable and plant raw materials
(Papchenko, Bologa, Berzoi, 1988; Grishko, Kozin, Chebanu, 1991),
food stuffs processing (Miyahara, 1985), winemaking (Kalmykova,
1993), and sugar production (Gulyi, Lebovka, Mank, Kupchik,
Bazhal, Matvienko et al., 1994; Jemai, 1997). But all these
electric field applications are usually restricted by the high and
uncontrolled increase in cellular tissue
temperature and
product quality deterioration because of electrode material
electrolytic reactions, etc.

Recently, a variety of new high and moderate pulsed electric field
(PEF) applications were successfully demonstrated for liquid and
solid foods (Barbosa-C\'{a}novas, Pothakamuri, Palou \& Swanson,
1998; Wouters \& Smelt, 1997; Knorr, Geulen, Grahl \& Sitzmann,
1994; Knorr \& Angersbach, 1998, Barsotti \& Cheftel, 1998).
The PEF application provides a possibility of fine regulation of
electric power input and may result in effective permeabilization
of cellular membranes (Zimmermann, 1975; Chang et al., 1992;
Weaver \& Chizmadzhev, 1996) without significant temperature
elevation (Barbosa-C\'{a}novas et al., 1998).

One of emerging and promising method is the combined PEF and
pressure application, which demonstrates significant yield
intensification for juice extracted from apples and beets and
clarification of the extracted juice (Vorobiev, Bazhal \& Lebovka,
2000; Gulyi et al., 1994). 
But the major problem arising
from simultaneous application of mechanical expression and PEF
treatment is the choice of optimal modes of treatment.
The mechanism of solid/liquid expression from cellular materials
is rather complex and may include many different phases of
consolidation process (Lanoiselle, Vorobiev, Bouvier \& Piar,
1996). The electric breakdown of a cellular system can influence
consolidation phases and change drastically the expression curves.
Unfortunately, up to now there are no accepted mechanism of
electric breakdown in the cellular systems and reliable criteria
for choosing optimal parameters of electric field treatment
(Lebovka et al., 1996, Lebovka, Bazhal \& Vorobiev, 2000). Another
problem is the poor reproducibility of the experimental data,
which is typical for objects of biological origin.

The properties of cellular materials influence significantly the
the electrotreatment efficiency.
The electrometry can be used for characterization of changes in
tissue properties under the influence of external factors
(electric field, pressure).
This is a simple method, as far as the electrical conductivity,
$\sigma$, reflects a degree of a water saturated tissue
permeability (Sahimi, 1995). But the general dependence between
the structure of a cellular material and $\sigma$ may be rather
complex, because the conductivity of a biological tissue may be
influenced by a number of processes, such as resealing of
membranes in cells (Heinz, Angersbach \& Knorr, 1999), diffusional
redistribution of moisture inside the samples, etc.

The objective of this study is the optimization of moisture
expression from biological raw materials under simultaneous
pressing and
PEF treatment. 

The liquid expression from fine-cut cellular tissue after PEF
application for different modes and durations of precompression
has been investigated. Apple was used as the example of cellular
material.

The useful method of data treatment, which allows to reduce data
scattering caused by differences in quality of the samples is
described. Discussion of the consolidation kinetics before and
after PEF treatment is also given.

\section{Preliminary remarks}

\label{Preliminary}


The reasons for the simultaneous application of mechanical
expression and PEF treatment are as follows. The excessive
quantity of extraparticle liquid and absence of contacts between
solid particles increases electrical energy losses. So the
effectiveness of the PEF treatment is restricted by the uniform
and tight packing of raw material between electrodes and previous
removing of extraparticle air and excessive liquid (from cells
destroyed by cutting). The method of raw material compact
formation is its pre-consolidation. Moreover, effectiveness of the
PEF treatment with  respect to the water-saturated cellular
materials is restricted by the low values of moisture content,
$W$. The PEF treatment is ineffective for a water-saturated system
(which is the case for the fine-cut apple raw material) because of
electric breakage and uncontrolled increase of current flow
through the system. The initial steps of consolidation remove an
excess liquid from the extracellular volume. Therefore, we can
expect increase of the PEF treatment efficiency after
pre-consolidation of the raw material.

\section{Materials and methods}
\label{Materials}

\subsection{Preparation of apple slices}
\label{Preparation}

Freshly harvested apples of Golden Delicious variety were selected
for investigation and stored at $4{{}^{\circ }}$C until required.
The moisture content of apples $W$ was within 80-85\%. The
fine-cut apple pieces (3-5 mm diameter) were prepared from an
apple pap using rasp.

\subsection{Experimental setup and instrumentations}
\label{Experimental}

Figure \ref{f1} is a schematic representation of the experimental
set-up. All experiments were carried out using laboratory
filter-press cell equipped with an electrical treatment system.
The polypropylene frame had a cylindrical cavity compartment (20
mm thick, 56 mm in diameter). The cavity compartment of frame was
initially filled up with apple slices and then was tightly closed
from both sides by the steel plates. One of the plates, covered by
a filter cloth, was used as a stationary electrode. The other
plate was attached with an elastic rubber diaphragm. A mobile wire
gauze electrode was installed between the diaphragm and the layer
of apple slices. Pressure was applied to the layer of apple slices
through the elastic diaphragm using the hydraulic pressure
controller GDS 'Standard'  (GDS Instruments Ltd, UK) with water as
a pressure fluid. The pressure controller provided a constant
pressure from 1 to 30 bars. The yield of liquid was controlled by
balance PT610 (Sartorius AG, Germany).

A high voltage pulse generator, $1500$V-$15$A (Service
Electronique UTC, France) provided the monopolar pulses of
rectangular shape and allowed pulse duration $t_{i}$ varied within
the interval of $10-1000$ $\mu $s (to precision $\pm 2$ $\mu $s),
pulse repetition time $\Delta t$ within the interval of $1-100$ ms
(to precision of $\pm 0.1$ ms) and number of pulses $n$ within the
interval of $1-100000$. The conductivities were measured by
contacting electrode method with an LCR Meter HP $4284$A (Hewlett
Packard, 38 mm guarded/guard Electrode-A HP 16451B) for thin apple
slice samples at the frequency of $100$ Hz and with Conductimetre
HI$8820$N (Hanna Instruments, Portugal) for the apple juice
samples at a frequency of $1000$ Hz (these frequencies were
selected as optimal in order to remove the influence of polarising
effect on electrodes and inside the samples). Pulse protocols and
all the output data (current, voltage, impedance, pressure, juice
yield and temperature) were controlled using a data logger and
special software HPVEE v.4.01 (Hewlett-Packard) adapted by Service
Electronique UTC, France).

High resolution scanning electron microscopy (SEM) images were
obtained using the instrument XL30 ESEM-FEG (Philips, V=15 kV,
P=3.5 Torr). The "WET" chamber mode allowing observation of
hydrated apple specimens in their natural state was applied.

The optical absorbance of an expressed liquid was measured with
Photocolorimeter CO75 (WPA Ltd, UK) at the wavelength 520 nm. The
characteristic absorption spectra were determined with respect to
distilled water. Transmittance  of an expressed liquid was calculated
as a ratio of filtered and nonfiltered liquid absorptions. The liquid was
filtrated using a Whatman 2V filter paper.

\subsection{Methods}
\label{Methods}

All experiments were done using electric field voltages $U$ from
$200$ to $1500$ V, pulse duration $t_{i} = 100$ s, pulse
repetition time $\Delta t = 10$ ms, number of pulses $N = 50$,
constant pressure $P = 3$ bars and total time of mechanical
expression $t$ up to $10^{4}$ s. Pressure value of 3 bars was
accepted as the most efficient for exhibition of the effect of
simultaneous pressing and PEF treatment (Vorobiev et al., 2000).
The experiments were repeated, at least, five times.



\section{Results and discussion}
\label{Results}

\subsection{Phases of consolidation}
\label{Phases}

Figure \ref{f2} presents typical experimental curves of liquid
yield $Y_{r}$, instantaneous flow rate $v_{r}$, pressure $P_{r}$
and specific electrical conductivity $\sigma_{r}$ vs. time $t$.
For convenience of presentation, here all properties are reduced
to their maximal values, e.g., $Y_{r}=Y/Y_{max}$, etc., and flow
rate is determined as $v=dY/dt$.

Initially we observe a rather rapid
increase of liquid yield $Y$, and decrease of electrical
conductivity, $\sigma$. This behaviour corresponds to the layer
pre-compaction (at a constant velocity of elastic diaphragm
displacement)  and
expulsion of extraparticle air-liquid mixture. The maximum of
instantaneous flow rate $v$ is observed approximately at
$t=t_{v_{max}}=50-60$ s. In the absence of PEF treatment ($U=0$,
or $E=0$ ) the curve of $\sigma(t)$ temporarily stabilises in time
interval $300<t<1000$ s and by this moment the pressure $P$
reaches its maximal value of 3 bars. This behaviour correspond to
the end of pre-compaction period. Then, at $t>1000$, the
$\sigma(t)$ curve ($E = 0$) slightly rises, which can be explained
by mechanical rupture of residual cells, by the material
deterioration and by the effects of the biological activity of
microorganisms.

Typical SEM micrographs of apple tissue structure before and after
pressing are presented in Figs. \ref{f3}(a) and (b), 
respectively. The mean size of undamaged cells is of order
$100-200$ $ \mu$m. We see, that after pressing some of the cells
are destroyed, but there exist also intact cells. So, for a given
mode of treatment ($P=3$ bars, $t=600$ s) the cellular structure
is not completely disrupted after pressing and there exists some
isolated cells, which remain intact during pressing period. These
cells can be damaged or partially permeabilized by another
methods, for example, by PEF, thermal or another mode of
treatment.


In analysing the results presented in Figs. \ref{f2},
\ref{f3}(a,b) we can discern the following phases of press-cake
layer consolidation process:

\begin{itemize}

\item Phase I. Initial compaction of press-cake and expulsion
of air-liquid mixture, or pre-consolidation period
($0<t\lessapprox 2t_{v_{max}}\cong 100-120$). During this phase
the velocity of elastic diaphragm displacement is constant and a
maximum of liquid flow rate is observed.

\item Phase II. Mechanical rupture of cells and expulsion of liquid from
ruptured cells ($100-120 \lessapprox t\lessapprox 300-400$). The
decrease of both liquid flow rate and velocity of elastic diaphragm
displacement and
acceleration of the pressure increase are observed.

\item Phase III. Final consolidation of press-cake at constant
pressure, packing of press-cake and retardation of liquid flow rate
($100-120 \lessapprox t$). The liquid flowing from intracellular,
extracellular and extraparticle volumes is expressed from the
press-cake.
At the beginning of this phase a minimum value of the specific electrical conductivity is
observed. Moisture occupies all channels and so
the press-cake is said to be in an impregnated state.

\end{itemize}

These phases are shown schematically at the top of Fig. \ref{f2}.

\subsection{Methods of data analysis for combined pressing and PEF treatment}
\label{PEF}

Significant changes in kinetics of moisture expression and press-cake
consolidation can be observed after PEF treatment. Figure
\ref{f4}(a) presents some examples of experimental curves of liquid
yield $Y(\%)$ versus time $t$. It can be seen that $Y(t)$ curves
rise significantly after PEF application (applied in this case at
$t=600 $ s ) as the result of damage or partial permeabilization
of intact cells and subsequent expression of liquid.

But here, the main problem is in poor reproducibility of the
experimental $Y(t)$ data. The measured  curves of $Y(t)$ can
deviate substantially because of differences in initial humidity
of samples. This difficulty can be overcome by consideration of
the normalized or reduced liquid yield. This normalization
procedure was executed in two steps. We began with the first
normalized form of liquid yield defined as (see Fig. \ref{f4}(b)):
\begin{equation}
Y_{E}^{I}=Y_{E}(t)/Y_{E}(t_{max}),  \label{e1}
\end{equation}
where $t_{max}$ is the maximal time of pressing (here we use the
value $t_{max} = 5400$ s), $Y_{E}$ values correspond to the values
of $Y$ at different electric field strengths $E$.

We assume that all liquid yield curves should be equal in the time
range of $t<t_{p}$ for identical conditions of pressing. For
experiments with PEF application $(E\neq 0)$, we take the curve
$Y_{E=0}^{I}(t)$ for $E = 0$ as a reference. Then we calculate the
area under this curve for the time period $t<t_{p}$ and
renormalize all the curves $Y_{E}^{I} (t)$  so as to obtain the
same values of
\begin{equation} S_{E} =\int\limits_{0}^{t_{p}
}Y_{E}^{I} (t) dt,  \label{e2}
\end{equation}
for all curves. 

Normalization coefficient is given by
\begin{equation}
k_{E} =S_{E=0} /S_{E}.  \label{e3}
\end{equation}

The second normalized form of liquid yield curve is defined as
follows (see Fig. \ref{f4}(c)):
\begin{equation}
Y_{E}^{II} (t)=k_{E} \cdot Y_{E}^{I} (t).  \label{e4}
\end{equation}

The degree of intensification caused by PEF treatment, $Y^{*}$ can
be determined as the following ratio of the second normalized
forms for pressing with and without PEF treatment:
\begin{equation}
Y^{*}=Y_{E}^{II}(t_{max})/Y_{E=0}^{II}(t_{max}).  \label{e5}
\end{equation}
Here the values of $Y_{E,E=0}^{II}(t_{max})$ were determined in
the point of the maximal time of pressing $t=t_{max}$.

Another interesting property of the expressing is the mean
characteristic time, which characterizes durability of the
process. The theory developed for mechanical expression of
cellular materials (Lanoiselle et al., 1996) describes
expression process with the set of characteristic times for the
different expression phases. These are very valuable
characteristics of the expression processes but, in  practice, it
is very difficult to find them proceeding from the expression
curves. The main problem here is that we can never determine the
exact value of limiting expression quantity $Y_{\infty }$ at
$t\rightarrow \infty.$  For vegetable 
stuff  we are faced 
with the problem of high enzymatic destruction at continuous
expression. So we always stop the expression process at rather
long, but finite time (in our case we choose $t_{max}=5400$ s) and
the obtained values of $Y(t_{max})$ are of course less then the
actual values of $Y_{\infty}$. More general approach implies
evaluation of the first Moment of the function
$F(t)=Y(t_{max})-Y(t)$:
\begin{equation}
M^{1}=\frac{\int\limits_{0}^{t_{\max
}}F(t)tdt}{\int\limits_{0}^{t_{\max }}F(t)dt}
 \label{e6}
\end{equation}

It is easy to show, that for the simple exponential function
$F(t)=\exp (-t/\tau )$ and \ $t_{\max }\rightarrow \infty $ the
first Moment is equal
to the characteristic time, $M^{1}=\tau .$ For the finite but large values of $%
t_{\max },$ we have $M^{1}\approx \tau (1-\frac{\left( t_{\max
}/\tau \right) ^{2}}{2e^{t_{\max }/\tau }})$, and at $t_{\max
}/\tau =5$ \ the first Moment  equals to $M^{1}\approx 0.92\tau $.
So, in our case the value of $M^{1}$ may serve for approximate
estimation of the effective characteristic time constant of the
whole expression process.

It is useful to use this approach for crude estimation of
characteristic time or durability of expression process after PEF
intervention. In such a case we can treat $M^{1}$ as a mean
characteristic time of liquid expulsion processes reflecting the
summarized effect of all the mechanisms in a system. We can define
the coefficient of PEF-enhanced durability as
\begin{equation}
\tau^{*}=M^{1}_{E}/M^{1}_{E=0}.  \label{e7}
\end{equation}

This coefficient shows the degree of the durability increase after
the PEF treatment.

\subsection{Influence of field strength $E$ and time $t_{p}$ of PEF treatment}
\label{Influence}

\subsubsection{Liquid yield kinetics}
\label{Yield}

Figure \ref{f5} presents the curves of excess normalized liquid
yield $\Delta Y=Y_{E}^{II}-Y_{E=0}^{II}$  versus time $t$ at
different times of PEF application, $t_{p}$, and different values
of electric field strength, $E$. We have applied PEF treatment in
different characteristic moments:

\begin{itemize}
\item $t_{p}=0$ s (before pressing);
\item $t_{p}=20$ s (initial phase of pressing);
\item $t_{p}=120$ s (second consolidation phase and specific
conductivity of a tissue is low);
\item $t_{p}=330$ s (press-cake pressure is achieved a constant
value);
\item $t_{p}=600$ s and $t_{p}=5400$ s (final consolidation phase).
\end{itemize}

The range of voltage values used corresponds to conditions of
steady PEF-application regime without any disruption of electrical
treatment caused by overflow of acceptable maximal current value.
For the given pulses protocol, the steady electrical treatment
regime was observed for the voltages not exceeding the following
maximal values: $U_{max} = 600$ V for $t_{p} = 0$ s, $U_{max} =
1000$ V for $t_{p} = 600$ s, and $U_{max} = 1500$ V for $t_{p} =
5400$ s.

As can be seen from Fig. \ref{f5} the form of the liquid yield
$Y^{II}_{E}$ curves in the time interval of $t<300$ s practically
does not depend on the time of PEF input $t_{p}$ and the electric
field strength values $E$ (that corresponds to small values of
deviations $Y^{II}_{E\neq 0}-Y^{II}_{E=0}$). Only at longer time
intervals $t>300$ s in the phase III of consolidation, we can
observe a behaviour reflecting the mode of PEF treatment.

At short times of PEF treatment, $t_{p}\lesssim 100-200$ s, the
excess liquid yield increases rapidly with $E$ increase as compared
with untreated sample (see Fig. \ref{f5}(a-c)). But
ineffectiveness of PEF treatment at short $t_{p}$  is less in
comparison with the cases of later PEF application. It can be
explained by the influence of excess quantities of air and
extraparticle liquid in the cake pores. The sample is highly
water-saturated at earlier period of pre-consolidation and PEF
application during this period may cause dielectric breakage and
uncontrolled increase of current flow through the system. At such
conditions, the intensification degree of PEF treatment $Y^{*}$ is
rather low (see black square for $t_{p}=0$ in Fig. \ref{f6}(a)).

When we apply the PEF treatment later on, for example, at the
beginning of the phase III of consolidation,
at $t\thicksim 300-400$ s a liquid excess yield
seems to be less dependent on $E$ (Fig. \ref{f5}(d)).
 Pressure achieves its constant value at
this time and most of cells that could be destroyed mechanically
at a given pressure (3 bars in our experiments) are already
disrupted and most of liquid is evacuated from these cells.
The residual isolated intact cells are connected with electrodes
through the network of channels containing conductive moisture. At
such conditions, the transmembrane potential on intact cells
should be high enough, even at low values of $E$, and, therefore,
the effective electropermeabilization of cell can be attained even
at minimal values of electric field intensity ($E=170$ V cm$^{-1}$
in our experiments).

As we can see from comparison of typical micrographs of apple
tissue structure received for pressing with and without electrical
treatment (Figs. \ref{f3}(b) and (c)), PEF application at $t=330$
s ($E=500$ V cm$^{-1}$) causes almost complete destruction of the
material. The similar pictures are observed in the wide interval
of $E\thicksim 250-500$ V cm$^{-1}$ and micrographs allow us to
identify only certain quantity of single isolated intact cells.

Dependencies  of the coefficient of PEF enhanced durability $\tau
^{*}$ versus electric field strength, $E$ (Fig. \ref{f6}(b))
substantiate conclusions set forth above.
We see that the value of $\tau^{*}$ depends considerably  on the
electric field strength $E$ only at small time of PEF input
($t_{p}<100-200$ s). The best liquid excess yield as compared with
untreated sample may be obtained at the lowest applied field $E$
when the PEF is applied at an instant when the pressure in the
system reaches  a constant value ($t_{p}=300-400$ s).

\subsubsection{
Specific conductivity $\sigma$, flow rate $v$ and pressure $P$
kinetics}

\label{conductivity}

Figure \ref{f7} presents the experimental curves of (a) a specific
electrical conductivity, $\sigma$,  and (b) instantaneous flow
rate, $v$,  versus time $t$ for the compressed layer of apple
slices. The PEF treatment was applied at $t=t_{p}=120$ s at
different external field strengths, $E$. As can be seen from Fig.
\ref{f7}(a), the $\sigma(t)$ values begin to rise abruptly after
PEF treatment, and this behaviour becomes more pronounced  with
increasing $E$. Such rise of $\sigma(t)$ values corresponds to the
combined effect of mechanical rupture of cells and their
electrical permeabilization.

We present schematically  the model of PEF treatment with and
without pressing in Fig. \ref{f8}.
In the absence of mechanical pressure 
the effect of the hidden (or passive) electrical breakdown can be
rather important. The electrical conductivity of the whole system
depends not only on the destruction degree of individual cells,
but also on the character of their connectivity and the presence
of continuously conducting channels.

At first, after PEF treatment the real degree of electrical
breakdown is hidden and does not affect the conductivity of the
whole system. PEF treatment permeabilizates cell membranes and
intensifies diffusion processes. The time of electric conductivity
build-up in a cellular material after PEF treatment can be
estimated from the time constant of diffusion processes:
$\tau_{D}\sim d^{2}/6D\approx 1$ s, where $d\approx 100$ is a mean
cell dimension, and $D\approx 10^{-9}$ m$^{2}$ s$^{-1}$ is a
diffusion coefficient of an endocellular fluid. In general, this
effect can exhibit a wide distribution of time constants
$\tau_{D}$ because of differences in cell dimensions, diffusivity
of intracellular solutions,  degree of cell permeabilization, etc.
Moreover, it can be masked by another related phenomena of
resealing process. Heinz et al. (1999) observed that the
insulating properties of the cell membranes can be recovered
within a few seconds after the small power PEF treatment and it
results in decreasing of a cellular system conductivity.

Simultaneous effect of pressing and PEF treatment($P\neq 0$ and
$E\neq 0$) can cause the primary changes. First of all, mechanical
and electrical stresses can be coupled to cause membrane breakdown
in cells (Akinlaja \& Frederick, 1998). Then, the external
pressure enhances flowing of the fluid from destroyed cells to
extracellular and extraparticle channels. All these decrease the
retardation time of electric conductivity build-up after PEF
treatment.
Under simultaneous PEF treatment and material compression
we can also eliminate or diminish the effect of a hidden electric
breakdown and to depress the cells resealing processes. So, in a
general case, we should observe in final state $\sigma_{P\neq
0}>\sigma_{P=0}$. This conclusion is confirmed by the data on
$\sigma(t)$ kinetics at $P=0$ presented in Fig. \ref{f7}(a) by the
dashed line. In this experiment we have dropped the pressure after
PEF treatment. This results in considerable decrease of
$\sigma(t)$ values as compared with the curve obtained at $P\neq
0$ (Fig. \ref{f8}).

A liquid flow rate $v$ also depends upon the degree of cellular
system destruction as a result of PEF treatment. But the behaviour
of $v(t)$ after PEF application does not change drastically (Fig.
\ref{f7}(b)). We can explain such behaviour by the fact of
substantial decrease and termination of liquid expulsion from the
press-cake by that time. The most pronounced peak is observed only
at the highest electric field strength ($E=400$ V cm$^{-1}$, Fig.
\ref{f7}(b)), and behaviour of the $v(t)$ curve reveals only
increase of its long time tails with increase of $E$ .

Figure \ref{f9} presents the experimental curves of (a) reduced
pressure values $P^{*}=P/P_{max}$  and (b) pressure difference
$P_{0}-P_{E}$  versus time $t$ for different durations of
pre-compression stage before the 
PEF application. It can be seen that PEF treatment diminishes
tissue rigidity which corresponds to the decrease of an effective
pressure in a system. Electrical treatment during the pressing
period up to $330$ s delays pressure increase and accelerates the
liquid yield. PEF application ($t_{p}>330$ s) can decrease
steady-state pressure abruptly and initiate a rise of a liquid flow
rate. The peaks of pressure decrease become sharper with
pre-compression time increase. It corresponds to the more rapid
destruction of cellular material pre-compressed during a longer
time. But then, at the final stage, the mechanical properties of a
press-cake reflate and pressure increases again up to the maximal
level.

\subsection{Optical properties of expressed liquid}

The expressed liquid (apple juice) absorption (or coloration) and transmittance change in
the course of a simple pressing process, due to filtration
properties of the cellular material in the course of time. In the final phase
of consolidation (phase III, see Fig. \ref{f2}), the liquid
coloration reduces considerable as compared with coloration of an
initial portion of moisture. As we have demonstrated before
(Vorobiev, Bazhal \& Lebovka, 2000), the simultaneous pressing and
electric field treatment result in considerable reduction of
coloration of those differential portions of  liquid which were
obtained after PEF application. Studying the temporal dependencies
of optical properties in differential liquid for different modes of
pressing and PEF application is {\it per se} of great interest.

Here we will discuss only the optical properties of the cumulative
expressed liquid, which is obtained as a result of combined pressing and PEF
treatment at the final stage of the process. Figure \ref{f10}
presents dependencies of absorption and transmittance versus
electric strength $E$ for extracted liquid at different values of
$t_{p}$. On the one hand, the PEF treatment decreases the liquid
coloration, as can be seen from absorption curves in
Fig.\ref{f10}. This is a positive factor of PEF treatment. We can
explain this phenomena by improvement of the tissue filtration
properties during pressing. Moreover, filtration properties of the
PEF treated press-cake also get improved with $t_{p}$ increase
because of increase of the pressed tissue consolidation. On the
other hand, transmittance of expressed liquid reduces with $t_{p}$ decrease
and increase of $E$. This is an undesirable phenomena. It can be
explained by the influence of electrical treatment on the
press-cake filtration properties. It is known that PEF application
causes many defects of the tissue. Application of the PEF
treatment increases yield of a liquid with high contents of
suspended particles.

That's why it is so important to choose a proper instant for PEF
application allowing to obtain the cumulative liquid with low
coloration and high transmittance. The PEF application at a moment
when pressure in the system achieves a constant value is
consistent with requirement of the best quality of a juice.

\section{Conclusion}

Investigations of the moisture expression from fine-cut apple raw
material under simultaneous mechanical expression and PEF
treatment were done. All experiments were performed using both
laboratory filter-press cell and high voltage pulse generator
which provided monopolar pulses of rectangular shape. The PEF
treatment was applied to materials that were expressed at time $t
= t_{p}$. Then the yield of liquid was analysed in comparison with
that of untreated material. The experimental results were obtained
for electric field strength $E$ varying from $200$ to $1500$ V
cm$^{-1}$, pulse duration $t_{i}$ = 100 s, pulse repetition time
$\Delta t = 10$ ms, number of pulses $N = 50$, constant pressure
$P = 3$ bars and maximal time of mechanical expression
$t_{max}=5.4\times10^{3}-10^{4}$ s.

The summary of results is as follows:

(1) The data obtained allows us to conclude that all kinetics
curves ( $\sigma(t)$, $Y(t)$,  $v(t)$ and $P(t)$) clearly reflect
three main consolidation phases in cellular material.

(2) The combination of pressing and PEF treatment gives the most
optimum results and permits to enhance significantly the liquid
yield in comparison with samples untreated by PEF. The PEF
treatment application permits to intensify pressing process
whenever the PEF is applied. But best liquid excess yield results
at lowest value of applied field $E$ may be obtained when PEF is
applied after some pre-compression period. Such pressure
pretreatment before PEF application is necessary for structuring
uniformity of the press-cake, removing excess moisture and
decreasing the electrical conductivity of cellular material. In our study,
the pre-compression period duration of $300-400$ s was optimal and
for which  the pressure in the system reaches a constant value.
The PEF application in this moment of time results in the best
quality of the expressed liquid (apple juice), which is confirmed by
its coloration and transmittance.

(3) The simultaneous pressure and PEF treatment application
reveals the passive form of the electrical damage. Electrical
damage under a low field without pressure application develops
very slow. The pressure provokes damage of defected cells,
enhances diffusion migration of moisture and depresses cells
resealing processes.

(4) The  proposed unified approach for liquid yield data analysis
allows to reduce the data scattering caused by the differences in
the quality of samples.

\section*{Acknowledgements}

The authors would like to thank the ``Pole Regional Genie des
Procedes`` (Picardie, France) and the Society CHOQUENET for
providing the financial support. Authors also thank Dr. N. S.
Pivovarova and Dr. A. B. Jemai for help with the preparation of
the manuscript.

\newpage

\section*{References}


Aguilera, J.M., \& Stanley, D.W. (1999). {\it Microstructural
principles of food processing and engineering}. Gaithersburg:
Aspen Publishers.

Akinlaja J., \& Frederick, S. (1998). The Breakdown of Cell
Membranes by Electrical and Mechanical Stress. {\it Biophysical
Journal, 75}(1), 247-254.

Barbosa-C\'{a}novas, G.V., \& Vega-Mercado, H. (1996). {\it
Dehydration of Foods} (pp. 289-320). New York: Chapman \& Hall.

Barbosa-C\'{a}novas, G.V., Pothakamury, U.R., Palou, E., \&
Swanson, B.G. (1998). {\it Nonthermal Preservation of Foods} (pp.
53-72). New York: Marcel Dekker.

Barsotti, L., \& Cheftel, J.C. (1998). Traitement des aliments par
champs electriques pulses. {\it Science des Aliments, 18}(6),
584-601.

Bazhal, M.I., \& Vorobiev, E.I. (2000). Electric treatment of
apple slices for intensifying juice pressing. {\it Journal of the
Science of Food and Agriculture} (in press).


Chang, D.C., B.M. Chassy, J.A. Saunders, \& Sowers, A.E., Eds.
(1992). {\it Guide to electroporation and electrofusion}. San
Diego: Academic Press.

Grishko, A.A., Kozin, V.M., \& Chebanu, V.G. (1991). {\it
Electroplasmolyzer for processing plant raw material}, US Patent
no. 4723483.

Gulyi, I.S., Lebovka, N.I., Mank, V.V., Kupchik, M.P., Bazhal,
M.I., Matvienko, A.B., \& Papchenko, A.Y.~ (1994). {\it Scientific
and practical principles of electrical treatment of food products
and materials}. Kiev: UkrINTEI (in Russian).

Heinz, V., Angersbach, A., \&  Knorr, D. (1999). High electric
field pulses and membrane permeabilization. In {\it Proceedings of
the European Conference on Emerging Food Science and Technology},
Tampere, Finland, 22-24 November  1999, 34.

Jemai, A.B. (1997). {\it Contribution a l'etude de l'effet d'un
traitement electrique sur les cossettes de betterave a sucre.
Incidence sur le procede d'extraction}. Th\`{e}se de Doctorat,
Universit\'{e} de Technologie de Compi\`{e}gne, Compi\`{e}gne,
France.

Jones, G.C. (1988), Cossette pretreatment and pressing, {\it
International Sugar Journal, 90}(1077), 157-167.

Kalmykova, I.S. (1993). {\it Application of electroplasmolysis for
intensification of phenols extracting from the grapes in the
technologies of red table wines and natural juice}. PhD Thesis,
Odessa Technological Institute of Food Industry, Odessa, Ukraine
(in Russian).

Knorr, D., \& Angersbach, A. (1998). Impact of high intensity
electric field pulses on plant membrane permeabilization. {\it
Trends in Food Science and Technology, 9}, 185-191.

Knorr, D., Geulen, M., Grahl, T., \& Sitzmann, W. (1994). Food
application of high electric field pulses. {\it Trends in Food
Science \& Technology, 5}, 71-75.

Lanoiselle, J.L., Vorobiev, E.I., Bouvier, J.M. \& Piar G. (1996).
Modeling of Solid/Liquid Expression for Cellular Materials, {\it
AIChE Journal, 42}, 2057-2068.

Lebovka, N.I., Mank, V.V., Bazhal, M.I., Kupchik, M.P. \& Gulyi,
I.S. (1996). Cascade model of thermal electrical breakdown of
inhomogeneous systems. {\it Surface Engineering and Applied
Electrochemistry, 1}, 29-33.

Lebovka, N.I., Bazhal, M.I. \&  Vorobiev E. (2000). Simulation and
experimental investigation of food material breakage using pulsed
electric field treatment. {\it Journal of Food Engineering, 44},
213-223.

McLellan, M.R., Kime, R.L., \& Lind, L.R. (1991).
Electroplasmolysis and other treatments to improve apple juice
yield. {\it Journal of Science Food Agriculture, 57}, 303-306.

Miyahara, K. (1985). {\it Methods and apparatus for producing
electrically processed food stuffs}, US Patent no. 4522834.

Papchenko, A.Y., Bologa, M.K. \& Berzoi, S.E. (1988). {\it
Apparatus for processing vegetable raw material}, US Patent no.
4787303.

Ponant, J., Foissac, S. \& Esnault, A. (1988). The alkaline
extraction of sugar beet. {\it Zuckerindustrie, 113}(8), 665-676.

Rao, M.A., \& Lund, D.B. (1986). Kinetics of thermal softening of
food - a review. {\it J. of Food Proc. and Pres., 10}, 311-329.

Sahimi, M. (1995). {\it Flow and transport in porous media and
fractured rock}. Weinheim: VCH.


Scheglov, Ju.A., Koval, N.P., Fuser, L.A., Zargarian, S.Y.,
Srimbov, A.A., Belik, V. G., Zharik, B.N., Papchenko, A.Y.,
Ryabinsky, F.G., \& Sergeev, A.S. (1988). {\it Electroplasmolyzer
for processing vegetable stock}, US Patent no. 4723483.

Schwartzberg, H.G., (1983) Expression-related properties. In M.
Peleg \& E.B. Bagley, Eds., {\it Physical properties of food} (pp.
423-472). AVI Pupl. Comp., Connecticut.

Tsuruta, T., Ishimoto, Y., \&  Masuoka T. (1998). Effect of
glycerol on intracellular ice formation and dehydration of onion
epidermis. {\it Annals of the New York Academy of Sciences, 858},
217-226.

Vorobiev, E.I., Bazhal, M.I., \& Lebovka, N.I. (2000).
Optimization of pulsed electric field treatment of apple slices by
pressing. In {\it Proceedings of the 8$^{th}$ International
Congress on Engineering and Food ICEF8}, Puebla, Mexico, 9-13
April, 2000, 265.

Weaver, J.C., \& Chizmadzev, Yu.A. (1996). Theory of
electroporation: A review. {\it Biolectrochemistry and
Bioenergetics, 41}(1), 135-160.

Zimmermann, U. (1975). Electrical breakdown:
electropermeabilization and electrofusion. {\it Reviews of
Physiology Biochemistry and Pharmacology, 105}, 176-256.



\begin{figure}[tbp]
\caption{Experimental set-up. } \label{f1}
\end{figure}

\begin{figure}[tbp]
\caption{ Typical experimental curves for different reduced
properties:  liquid yield $Y_{r}$, instantaneous flow rate $v_{r}$,
pressure $P_{r}$ and specific electrical conductivity $\sigma_{r}$
versus time of the expression $t$. For the convenience of
presentation all properties here are reduced to their maximal
values, e.g., $Y_{r}=Y/Y_{max}$, etc., flow rate is determined as
$v=dY/dt$. The error bars represent standard deviations of data.
At the top we show the proposed scheme of changes in the cellular
tissue 
structure with time.} \label{f2}
\end{figure}

\begin{figure}[tbp]
\caption{SEM micrographs of apple tissue: (a) before treatment,
(b) after pressing ($P=3$ bars, $t=600$ s), (c) after simultaneous
PEF \& pressing treatment ($P=3$ bars, $t_{p}=330$ s, $E=500$ V
cm$^{-1}$, $t=600$ s).} \label{f3}
\end{figure}

\begin{figure}[tbp]
\caption{ Execution steps in the procedure of experimental curves
normalization: examples of real experimental curves of expressed liquid
yield $Y(\%)$ versus time $t$ obtained for $E=520$ V cm$^{-1}$ and
$t_{p}=600$ s (a); definition of the  first normalized form of
liquid yield $Y_{E}^{I}$ (b); and definition of the second
normalized form of liquid yield $Y_{E}^{II}$ (c).} \label{f4}
\end{figure}

\begin{figure}[tbp]
\caption{ Kinetics of normalized excess liquid yield
$\Delta Y=Y_{E}^{II}-Y_{E=0}^{II}$ at
different values of $t_{p}$and $E$. The error bars represent
standard deviations in the data.}

\label{f5}
\end{figure}

\begin{figure}[tbp]
\caption{ Degree of PEF treatment intensification $Y^{*}$ (a) and
coefficient of PEF enhanced durability $\tau ^{*}$ (b) versus
electric field strength $E$ at different values of PEF treatment
application time $t_{p}$. The error bars represent standard
deviations in the data.}

\label{f6}
\end{figure}

\begin{figure}[tbp]
\caption{ Specific electrical conductivity, $\sigma$, (a) and
instantaneous flow rate $v$ (b) versus time of expression, $t$, at
$t_{p}=120$ s and different values of $E$. The error bars
represent standard deviations in the data.}

\label{f7}
\end{figure}

\begin{figure}[tbp]
\caption{ Schematic of the model of PEF treatment with and without
pressure $P$. The conductivity of the sample is higher for case
when the pressure is applied.}

\label{f8}
\end{figure}

\begin{figure}[tbp]
\caption{ Reduced pressure $P^{*}=P/P_{max}$ (a) and pressure
difference $P_{E=0}-P_{E}$ (a) versus time of expression, $t$, at
different values of   $t_{p}$ and $E$. The error bars represent
standard deviations in the data.}

\label{f9}
\end{figure}

\begin{figure}[tbp]
\caption{Absorbance and transmittance of liquid yielded during
pressing versus electric field strength, $E$, at different values
of PEF treatment application time, $t_{p}$. The error bars
represent standard deviations in the data.}

\label{f10}
\end{figure}

\end{document}